\documentclass{aa}

\usepackage{graphicx} 
\usepackage{natbib} 
\bibpunct{(}{)}{;}{a}{}{,}

\begin{document} 
   \title{Numerical simulation  of the three-dimensional structure 
     and dynamics of the non-magnetic solar chromosphere}
   \titlerunning{Numerical simulation of the non-magnetic solar chromosphere}

   \author{Sven Wedemeyer \inst{1,2} 
      \and Bernd Freytag \inst{3} 
      \and Matthias Steffen \inst{4} 
      \and Hans-G\"unter Ludwig \inst{5} 
      \and Hartmut Holweger \inst{1}} 
 
   \authorrunning{S.~Wedemeyer et al.} 
 
   \offprints{wedemeyer@kis.uni-freiburg.de} 
 
   \institute{Institut f\"{u}r Theoretische Physik und Astrophysik,  
     Universit\"{a}t Kiel, 24098~Kiel, Germany
     \and Kiepenheuer-Institut f\"{u}r Sonnenphysik, Sch\"{o}neckstrasse~6, 
     79104~Freiburg, Germany      
     \and Department for Astronomy and Space Physics, Uppsala University,  
     Box~515, 75120~Uppsala, Sweden 
     \and Astrophysikalisches Institut Potsdam, An der Sternwarte~16,  
     14482~Potsdam, Germany 
     \and Lund Observatory, Box 43, 22100~Lund, Sweden} 
 
   \date{Received date; accepted date}

\abstract{
Three-dimensional numerical simulations with
\mbox{\textsf{CO$^\mathsf{5}$BOLD}}, a new
radiation hydrodynamics code, result in a dynamic, thermally
bifurcated model of the non-magnetic chromosphere of the quiet Sun.
The \mbox{3-D}~model includes the middle and low chromosphere, the
photosphere, and the top of the convection zone, where acoustic waves
are excited by convective motions.  While the waves propagate upwards,
they steepen into shocks, dissipate, and deposit their mechanical
energy as heat in the chromosphere.
Our numerical simulations show for the first time a complex \mbox{3-D}
structure of the chromospheric layers, formed by the interaction of
shock waves.  
Horizontal temperature cross-sections of the model chromosphere
exhibit a network of hot filaments and enclosed cool regions.  
The horizontal pattern evolves on short time-scales of the order of typically 
\mbox{20 - 25} seconds, and has spatial scales comparable to those of the 
underlying granulation. 
The resulting thermal bifurcation, i.e., the co-existence of cold and
hot regions, provides temperatures high enough to produce the observed
chromospheric UV emission and -- at the same time -- temperatures cold
enough to allow the formation of molecules (e.g., carbon monoxide).
Our \mbox{3-D}~model corroborates the finding by \citet{carlsson94}
that the chromospheric temperature rise of semi-empirical models does
not necessarily imply an increase in the average gas temperature but
can be explained by the presence of substantial spatial and temporal
temperature inhomogeneities.
\keywords{Sun: chromosphere, hydrodynamics, radiative transfer} 
} 
\maketitle

\section{Introduction} 
\label{sec:intro} 

Three-dimensional, time-dependent radiation hydrodynamics
simulations of solar and stellar surface convection have now reached a
level of sophistication which goes far beyond that of idealised
numerical experiments, and allows a direct confrontation of such
models with real stars
\citep[e.g.,][]{stein98,asplund00,cobold,ludwig02}. Extending this
kind of simulation to include the low chromosphere, it is possible to
study \-- in a single model and based on first principles \-- the
generation of waves by the convective flow as well as the wave
propagation and dissipation in the higher layers. Extended simulations
of this type may then be utilised to explore the hitherto poorly
understood \mbox{3-D} thermal structure and dynamics of the
non-magnetic chromospheric internetwork regions, and to obtain an
independent theoretical estimate of the amount of chromospheric
heating due to acoustic waves.

A strong motivation for three-dimensional time-dependent modelling arises
from the need to reconcile apparently contradictory solar
observations: Carbon monoxide absorption lines imply gas temperatures
as low as $\approx 3700$~K in the chromosphere of the quiet Sun
\citep[see ][ and references therein]{
noyes72a,ayres81b,solanki94,uitenbroek94,uitenbroek00a,ayres02},
whereas chromospheric UV emission features require much higher
temperatures at the same heights \citep[e.g.,][]{ayres76,carlsson97b}.
 
Semi-empirical models which have been constructed based on UV and microwave 
observations (\citealp[e.g.,][ hereafter VAL]{val81};
\citealp[][]{maltby86}; \citealp[][ hereafter FAL]{fal93}) 
commonly feature a temperature minimum of $T_{\mathrm{min}} \approx
4200\,-\,4400$~K at a height of $z \approx\,500$~km above optical
depth unity and an outwardly increasing temperature above. On the
other hand, models based on CO observations
\citep[e.g.,][]{wiedemann94} show a monotonic decrease of temperature
with height.

These conflicting observations and the inferred representative models have
led to a controversy about the nature of the chromosphere of the
non-magnetic quiet Sun which is going on for many years now \citep[see, 
e.g.,][]{kalkofen01}: Is the chromosphere of the average quiet Sun a
time-dependent phenomenon with a mostly cool background and large
temperature fluctuations due to upward propagating shock waves? Or is
it persistent and always hot with only small temperature fluctuations?
In short: Is the non-magnetic solar chromosphere hot or cool?

A large number of observations show that the chromosphere of the quiet 
Sun is indeed a very dynamic phenomenon 
\citep[e.g.,][]{carlsson97b,muglach01, krijger01, wunnenberg02}. 
Obviously, static one-dimensional models can only describe selected 
time-averaged properties. 
More realistic modelling should therefore be time-dependent.

Starting in the late 1960s, the pioneering work on \mbox{1-D}
time-dependent numerical models of chromospheric heating by acoustic
and magneto-hydrodynamic waves is due to Ulmschneider and
collaborators.  In a long series of papers
\citep[e.g.,][]{ulmschneider71,ulmschneider77,ulmschneider78,muchmore85,
ulmschneider87,ulmschneider89,cuntz94},
they studied in detail the chromospheric energy balance between
dissipation of prescribed short-period (mostly monochromatic) acoustic
waves and radiative emission. In their models, the acoustic energy
flux is supplied by a piston acting as a lower boundary condition.
Assuming that the generation of acoustic waves by the ``turbulent'' flows in
the upper convection zone can be described by the Lighthill-Stein theory
\citep[][]{lighthill52,stein67,stein68,musielak94,ulmschneider96,
ulmschneider99}, they compute dynamic chromospheric models not only
for the Sun but also for a sample of main-sequence stars and
giants. Based on these models, they conclude that the observed ``basal
flux'' from the chromospheres of late-type stars
\citep[][]{schrijver87,rutten91} is fully attributable to the
dissipation of acoustic wave energy
\citep[][]{buchholz98}, and that the observed variation 
 of chromospheric emission can be explained by the additional heating of
 magnetohydrodynamic shock waves \citep{ulmschneider01}.

The detailed radiation hydrodynamics simulations by   
\citet[][ hereafter CS]{carlsson94,carlsson95,carlsson97a}
are another prominent example of sophisticated \mbox{1-D}
time-dependent modelling. These authors successfully explained the
\ion{Ca}{ii}~\mbox{H$_{\rm 2v}$} bright points as a result of
propagating shock waves. In their model, the waves are excited by a
piston which is driven by a velocity variation derived from observed
oscillations at the photospheric level.  Instead of a temperature
minimum and a monotonic temperature increase above, as characteristic
of the VAL and FAL models, CS find a chromosphere with a mostly cool
background and large temperature fluctuations due to upward
propagating shocks. Even more remarkable is the fact that they are
able to reproduce the rise of the radiation temperature without
an increase of the mean gas temperature. Basic reasons are the
nonlinear temperature dependence of the Planck function in the UV and
the extreme temperature peaks associated with the shock waves.  This
led CS to the conclusion that the chromosphere of the quiet Sun is not
persistent but a spatially and temporally intermittent phenomenon
which -- if averaged over space and time -- is mostly cool and not hot.

Although the one-dimensional models of the non-magnetic solar
chromosphere mentioned above are highly elaborate, including a
fully time-dependent \element{H} ionisation and detailed NLTE 
radiative transfer, they suffer from the need for an external
prescription of the wave excitation, and of course they cannot 
account for horizontal inhomogeneities 
and the associated effect of dynamic cooling on the atmospheric energy 
balance.
 
In this regard, the  three-dimensional self-consistent modelling by 
\citet{skartlien00c} can be considered as a major improvement. 
The idea of Skartlien and co-workers was to extend the standard
radiation hydrodynamics simulations of the solar granulation
\citep{stein98} into the chromosphere, where local
thermodynamic equilibrium (LTE) is known to be a
poor approximation. 
In order to adapt it to chromospheric conditions,
\citet[][]{skartlien00b} upgraded the radiative transfer part of the
Nordlund-Stein code by implementing an iterative method to treat
coherent isotropic scattering in \mbox{3-D}.  The simulations enabled
\citeauthor{skartlien00c} to analyse the generation, propagation, and
dissipation of acoustic waves in three dimensions.  The main emphasis
of their study was on the excitation of transient wave emission
resulting from the collapse of small granules, and the dynamic 
response of the chromospheric layers to such acoustic events. 
    
In the present paper, we present similar time-dependent
\mbox{3-D}~models which extend from the upper convection zone to the
middle chromosphere. The radiation hydrodynamics simulations are
performed with \mbox{\textsf{CO$^\mathsf{5}$BOLD}}, a new radiation hydrodynamics 
code developed by B.~Freytag and M.~Steffen \citep{cobold}.  In this
exploratory simulation, we treat the radiative transport in LTE with
grey opacities (see Sect.~\ref{sec:rad} and discussion in
Sect.~\ref{sec:discuss}). This simplification allows us to work at a
significantly higher spatial resolution ($140
\times 140 \times 200$ cells) than \citeauthor{skartlien00c} ($32
\times 32 \times 100$ grid). We find that the \mbox{3-D} structure of
the non-magnetic chromospheric layers is characterised by a complex
pattern of interacting shocks, forming a network of hot filaments and
enclosed cool ``bubbles''. 
This chromospheric pattern and its implications are chosen as major subject 
of this paper since the topology and the dynamics of the pattern are likely not 
to be too sensitive to the LTE simplification.
We conclude that the low chromosphere
exhibits a prominent thermal bifurcation: hot and cool regions exist
side by side.  Surprisingly, this small-scale (non-magnetic) network
was not mentioned by \citeauthor{skartlien00c}; presumably, it was not
noticed due to the poor (horizontal) spatial resolution of their
numerical model.

In Sect.~\ref{sec:cobold} we will give a short overview of the
numerical details of \mbox{\textsf{CO$^\mathsf{5}$BOLD}}.  The \mbox{3-D}~model is described in
Sect.~\ref{sec:model}, followed by the results in Sect.~\ref{sec:results}.
Finally, a discussion and conclusions are presented in
Sect.~\ref{sec:discuss} and Sect.~\ref{sec:conclusion}, respectively.

\section{The radiation hydrodynamics code \mbox{\textsf{CO$^\mathsf{5}$BOLD}}} 
\label{sec:cobold} 

\mbox{\textsf{CO$^\mathsf{5}$BOLD}}\ solves the time-dependent hydrodynamic equations coupled with the  
radiative transfer equation for a fully compressible, chemically homogeneous  
plasma in a constant gravitational field in 
two or three spatial dimensions.
Operator splitting separates Eulerian hydrodynamics, \mbox{3-D}  
tensor viscosity, and radiation transport. 
Magnetic fields are not included so far, restricting this version of \mbox{\textsf{CO$^\mathsf{5}$BOLD}}\   
to internetwork regions.

The most important properties of the code are described below  
\citep[see also][]{cobold,wedemeyerphd}.  
A more detailed paper on the code itself is in preparation (Freytag, in prep.).  

\begin{figure*}[t] 
\centering 
  \resizebox{\hsize}{!}{\includegraphics{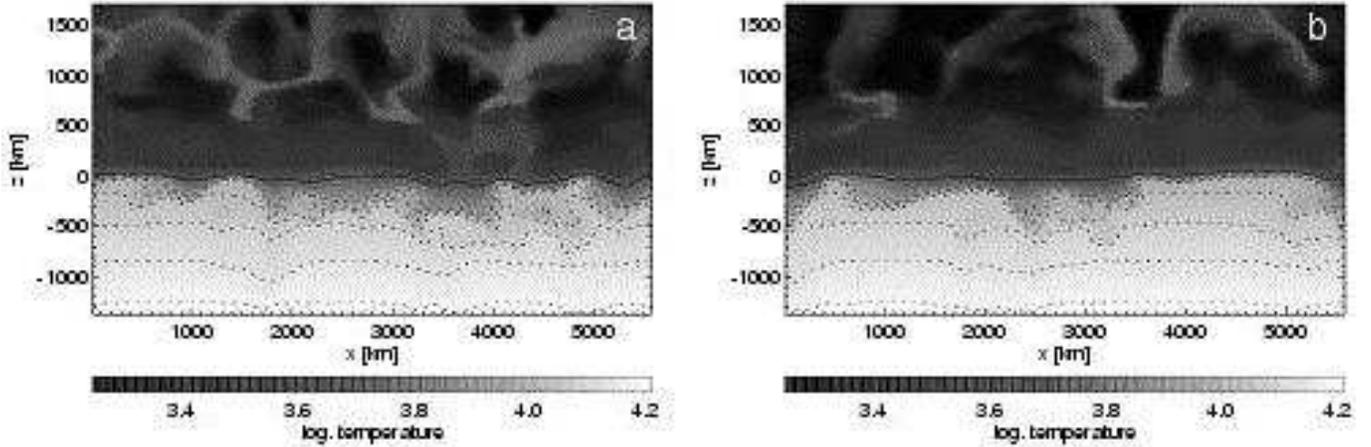}} 
  \caption{Logarithmic temperature in vertical \mbox{2-D} slices taken from the  
    \mbox{3-D}~model at different horizontal positions \mbox{(\,\,\textbf{a)} 
    $y = 1540$~km,} \mbox{\textbf{b)} $y = 2820$~km}~): Top  
    of the convection zone, photosphere, and low/middle chromosphere with  
    propagating shock waves. 
    The solid line marks the height for optical depth unity.  
    The dotted lines are contours for $\log T = 3.7, 3.95, 4.00, 4.05, .. ,4.20$  
    (top to bottom). 
    The temperature ranges from $\approx 16400$~K to  
    $\approx 5200$~K in the convection zone (i.e., below  
    $z = 0$~km) and decreases to $\approx 3000 $~K in the  
    photosphere and even down to $\approx 1800 $~K in the chromosphere. 
  } 
  \label{fig.xzslices} 
\end{figure*} 

\begin{figure*}[p] 
\centering 
  \includegraphics[width=17cm]{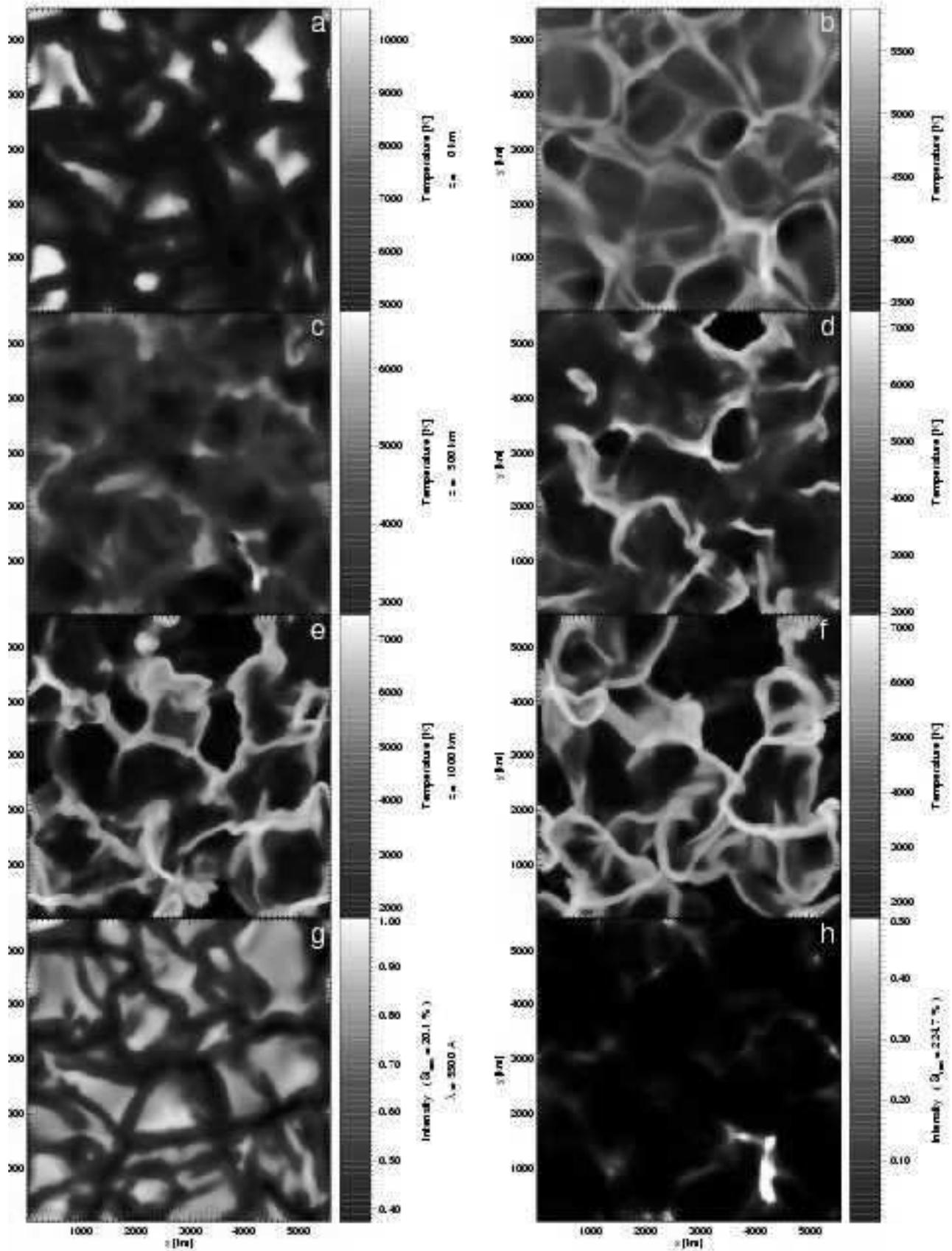}
  \caption{Temperature in horizontal \mbox{2-D} slices at different heights
     in the 
    photosphere at $z = 0$~km, $250$~km, and $500$~km \mbox{\textbf{(a-c)}}, and in the 
    chromosphere at $z = 750$~km,  
    $1000$~km, and $1250$~km \mbox{\textbf{(d-f)}}. 
   Panels \mbox{\textbf{g)}} and \mbox{\textbf{h)}} show the emergent continuum
   intensity at $\lambda = 5500$~\AA\ and $\lambda = 1600$~\AA, respectively. 
} 
  \label{fig.xyslices} 
\end{figure*} 

\begin{figure*}[t] 
\centering 
  \resizebox{\hsize}{!}{\includegraphics{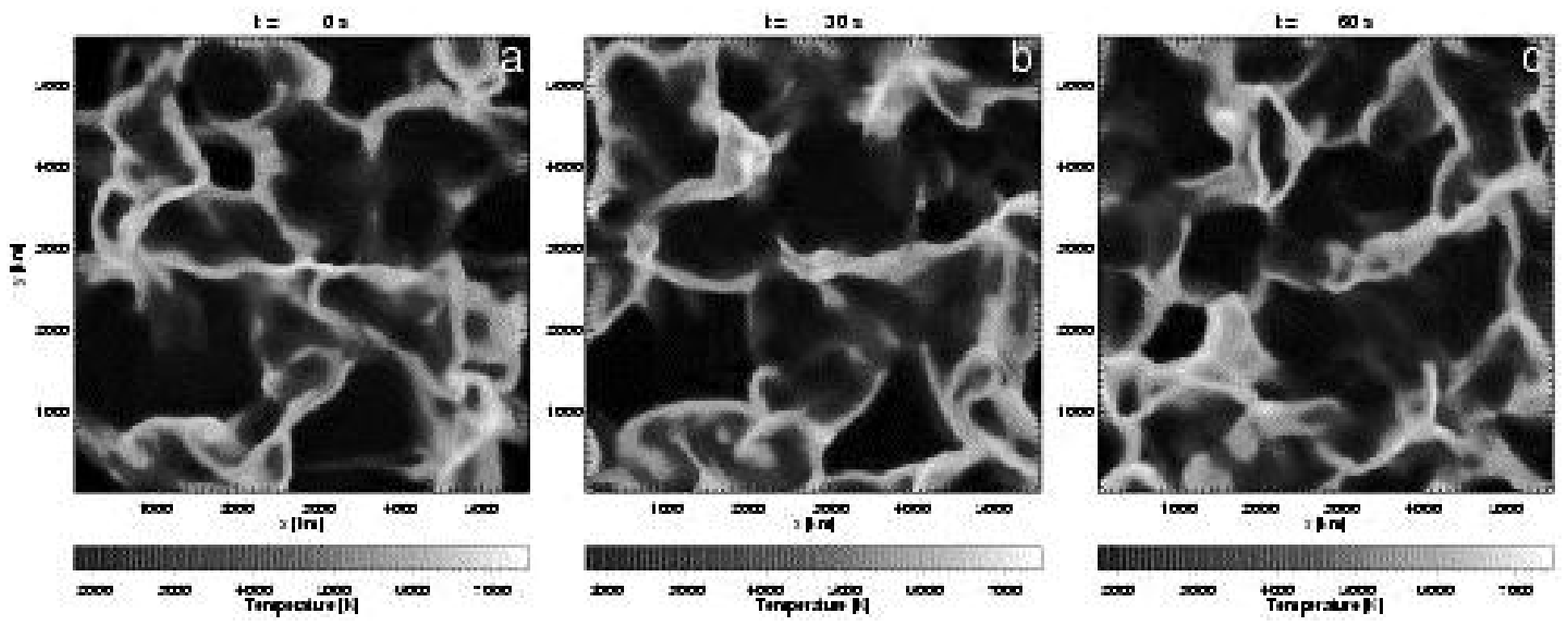}} 
  \caption{Temperature in horizontal \mbox{2-D} slices at  
    $z = 1000$~km for a short time sequence ($\Delta t = 30$~s).} 
  \label{fig.xytimeseq} 
\end{figure*}

\subsection{Hydrodynamics} 
\label{sec:hydro} 

The relations for the conservation of mass, momentum, and energy are
solved on a fixed Cartesian grid allowing spatially non-equidistant meshes.  
Directional operator splitting transforms the \mbox{2-D}/\mbox{3-D} problem into  
\mbox{1-D} sub steps which then can be treated with a fast approximate Riemann  
solver \citep{roe86}.  
The scheme is modified to account for a realistic equation of state and an  
external gravity field. 

Additionally, a small amount of tensor viscosity is added in a separate sub step.  
Although the hydrodynamics scheme is stable enough to handle \mbox{1-D} and most 
multi-dimensional problems, there are special multi-dimensional cases which  
require an additional tensor viscosity to ensure stability.  
Such cases occur, e.g., near strong shocks which are aligned with the grid  
\citep{quirk}. Our numerical scheme has proven to be very robust
in handling shocks, which is important when modelling chromospheric 
conditions.
 
\subsection{Radiation transport} 
\label{sec:rad} 

The equation of radiative transfer is solved applying long
characteristics (``rays'').  A large number of rays traverse the
computational box under different azimuthal and inclination angles.
Independently along each ray, the radiative transfer equation is
solved with a modified Feautrier scheme.  The radiation
transport is treated in strict LTE
so far.  In this work, a grey (frequency-independent) radiation
transport with realistic opacities is used (see Sect.~\ref{sec:eosopa}).
The applied scheme is well-suited for the lower layers (convection
zone and photosphere), but clearly requires further improvements for
chromospheric conditions where substantial deviations from LTE prevail
and the UV radiative transfer is dominated by scattering. 
See also the discussion in Sect.~\ref{sec:discuss}.

\subsection{Equation of state and opacities} 
\label{sec:eosopa} 

The equation of state takes into account partial ionisation of 
\element{H} and \element{He},
as well as formation and dissociation of $H_2$,
assuming thermodynamic equilibrium.
It is solved by interpolation in a table which is computed in advance
for a prescribed chemical composition of hydrogen, helium, and a
representative metal.
The table consists of two-dimensional arrays as functions of 
density and internal energy.

For the model presented in this work we used a Rosseland mean opacity look-up  
table which has been compiled and processed based on data of OPAL for  
temperatures above \mbox{$12000$~K} \citep{opal} and PHOENIX for  
temperatures below \mbox{$12000$~K}  
\citep[][ and references therein]{hauschildt97}.  
The table provides the opacity as a function of temperature and gas pressure.

Although a large number of atomic lines and molecular features are
formally taken into account in the construction of the opacity table,
it is clear that the stronger lines are not properly represented when
computing the grey opacity according to the Rosseland averaging
procedure. Consequently, the stronger spectral features are
essentially ignored in the present approach (see also Sect.~\ref{sec:discuss}).

\subsection{Boundary conditions} 
\label{sec:boundary} 

Located deep in the convectively unstable layers, 
the lower boundary is open, i.e., material is allowed to flow in and out of the  
computational box.  
The inflow of material is constrained to ensure a vanishing total mass flux  
across the lower boundary so that the total mass in the computational volume 
is preserved -- aside from smaller gains or losses across the upper boundary.  
The entropy of inflowing material is a prescribed parameter, and indirectly  
controls the effective temperature of a model.    
The vertical derivative of the velocity components is zero. 
The pressure in the bottom layer is kept close to plane-parallel by
artificially reducing horizontal pressure fluctuations towards zero
with a prescribed time constant.

At the upper transmitting boundary the vertical derivative of the
velocity components and of the internal energy are zero; the density
is assumed to decrease exponentially above the top boundary.
Material can flow into the computational box if the velocity at the
boundary is directed downwards.  The temperature of the inflowing
material is then altered towards a temperature $T_{\rm top}$ on a
characteristic time scale of typically a few seconds. 
This simple boundary condition turns out
to be stable and allows (shock) waves to leave the computational box
without noticeable reflections. Moreover, we have chosen the location
of the upper boundary such that it is far away from the regions which
are of particular interest in this work.

The lateral boundary conditions are periodic.

\section{The \mbox{3-D} model} 
\label{sec:model} 

The \mbox{3-D}~model consists of horizontally $140$ grid points ($x,y$) with a  
constant resolution of $40$~km, leading to a horizontal size of  
$5600$~km which corresponds to an angle of $\approx 7".7$ in  
ground-based observations.   
The total vertical height is $3110$~km, reaching from
the upper convection zone at \mbox{$z = -1400$~km} to the middle  
chromosphere at a height of $z = 1710$~km.  
The origin of the geometric height scale (\mbox{$z = 0$~km}) corresponds  
to the temporally and horizontally averaged Rosseland optical depth unity.  
In the following we refer to the photosphere always as the layer between  
$0$~km and $500$~km in model coordinates, and to the  
chromosphere as the layer above.  
The $200$ vertical grid points are non-equidistant, with a resolution of  
$46$~km at the bottom  which decreases with height down to a constant  
distance of $12$~km for all layers above $z = -270$~km.  
The computational time step is typically $0.1$ to $0.2$~s.
 
As an initial model, we extended an already evolved model which reached up 
to the top of the photosphere.  
The temperature and density stratification for the new grid cells were  
calculated under the assumption of hydrostatic equilibrium.  
Interestingly, the further evolution of the model does not depend strongly on 
the initial condition because the chromosphere turns out to be highly dynamical  
on short time-scales.  
After only a few minutes of simulation time the initial chromosphere already  
formed the typical structures which we will discuss below.  
 
However, the first $170$~min of the simulation sequence are not used for data  
analysis to ensure that the model has sufficiently relaxed.  
The results presented in this work are based on another $151$~min of 
simulation time.

\section{Results} 
\label{sec:results} 

\subsection{Structure of the model atmosphere} 
\label{sec:struct} 

\mbox{Figures~\ref{fig.xzslices}-\ref{fig.xytimeseq}} show the temperature in  
vertical and horizontal slices of the \mbox{3-D}~model which is described in 
Sect.~\ref{sec:model}. 
The data for Figs.~\ref{fig.xzslices}-\ref{fig.xyslices} are taken from the same 
time step. 
Figure~\ref{fig.xyslices}  illustrates the depth-dependence of the structure of the model atmosphere by 
means of \mbox{2-D} temperature slices at various geometrical heights. 
The same figure also shows synthetic images of the emergent continuum 
intensity at $\lambda = 5500$~\AA\ and $\lambda = 1600$~\AA\ which were 
computed subsequent to the simulation for the selected 
time step. 
For these calculations  LTE radiative transfer was assumed. 
We used pure continuum opacities (dominated by \ion{Si}{i} b-f 
absorption at 1600~\AA) taken from the Kiel spectrum synthesis package LINFOR. 

Obviously, there are striking differences between the horizontal patterns in 
the photosphere and the layers above. 
The temperature at the bottom of the photosphere \mbox{(Fig.~\ref{fig.xyslices}.a)} 
reveals the granulation which comes out more clearly in the intensity image for  
$\lambda = 5500$~\AA\ \mbox{(Fig.~\ref{fig.xyslices}.g)}. 
The granulation is very similar to observations in various aspects 
like shape, size distribution, and lifetime of the granules, indicating 
that in the lower part of the model the physics are realistically
represented \citep{wedemeyerphd}. 
Only $250$~km above, a reversed granulation pattern appears 
\mbox{(Fig.~\ref{fig.xyslices}.b)}:
the inner parts of the granules are dark due to the rapid cooling of
the ascending gas, and bright rims (note the double structure) appear 
at the edges of the granules, representing hot shocked gas being
directed into the intergranular lanes.

Higher up, the model chromosphere is characterised by a network of
hot matter and small-scale hot spots on a cool background as can be
seen in the horizontal cross-sections in
\mbox{Fig.~\ref{fig.xyslices}.d-f}. The pattern is a result of
interaction of propagating hydrodynamic shock waves which are an
ubiquitous phenomenon in the model chromosphere. The shock fronts are
usually inclined, so a horizontal cut through the temperature field
shows a filamentary structure. There is also a clear signature of
oscillations with periods in the 3-min range (see
Fig.~\ref{fig.oscheight}). Shock waves are present at all time steps,
mostly several at the same time
\mbox{(Figs.~\ref{fig.xzslices}-\ref{fig.xytimeseq})}.  The waves
propagate in the vertical as well as in the horizontal direction and
interfere with each other, compressing and heating the gas in the
filaments (see Sect.~\ref{sec:shockwaves}).

As a consequence of the correlation between convective motions and the  
excitation of acoustic waves, the spatial scale of the pattern is comparable to  
that of the underlying granulation.   

The network-like pattern appears more subtle in the UV continuum intensity at a 
wavelength of \mbox{$\lambda = 1600$~\AA}\  (\mbox{Fig.~\ref{fig.xyslices}.h}).
Rather, a small area of enhanced emission stands out of an otherwise dark 
background. 
This is caused by the highly non-linear temperature response of the
Planck function in the UV.
Hence, the hot gas, which is connected to the propagating shock waves, 
contributes by far more to the emergent UV continuum intensity 
than the cool regions. 
Note that for more realistic results, scattering and line blocking must be taken into account. 

Due to the ongoing propagation of the waves the pattern changes 
continuously (see Fig.~\ref{fig.xytimeseq}) on time-scales which are 
much shorter than derived for the granulation. 
We calculated autocorrelation times for sequences of horizontal temperature slices 
and determined height-dependent pattern evolution time scales as the time lags for 
which  the autocorrelation decreased to a value of $1/e$. 
At chromospheric heights the characteristic time scales are as short as $20 -25$~s
whereas the same analysis produces time scales of $\ga 120$~s at the bottom of 
the photosphere ($z=0$). 
Using the emergent grey intensity, which renders the low photosphere, 
instead of the gas temperature leads to $\sim 200$~s. 
The difference between temperature and intensity result can be understood if one 
considers that structures also move up and down, for instance, due to 
oscillations. 
Consequently, the pattern at a fixed geometrical height changes more quickly than 
visible in the corresponding intensity. 
Furthermore, spatial smearing of the pattern, i.e., reducing the image resolution 
to values caused by observational seeing conditions, produces longer time scales. 
This should be kept in mind when comparing the theoretical results with 
empirical data.

\subsection{Waves, oscillations, and shocks} 
\label{sec:shockwaves} 

\begin{figure}[t] 
  \resizebox{\hsize}{!}{\includegraphics{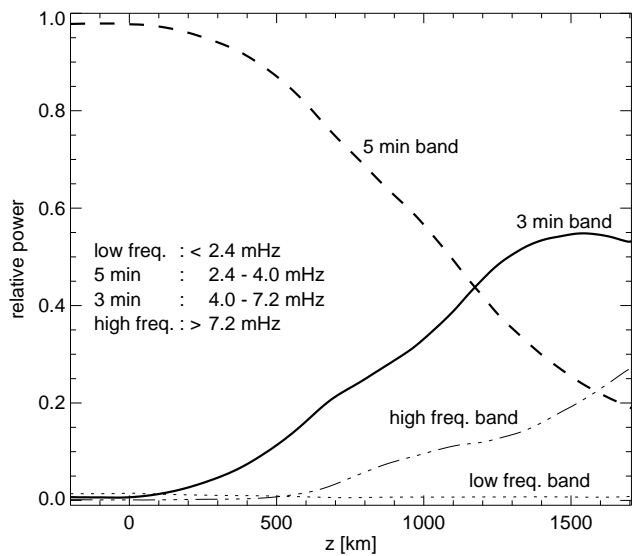}}
  \caption{Variation of relative velocity power with
  height. At each height the power spectrum of the horizontal average
  of the vertical
  velocity was integrated over the frequency intervals which are
  specified on the left.}
  \label{fig.oscheight} 
\end{figure} 

\begin{figure*}[t] 
  \centering 
  \includegraphics[width=12cm]{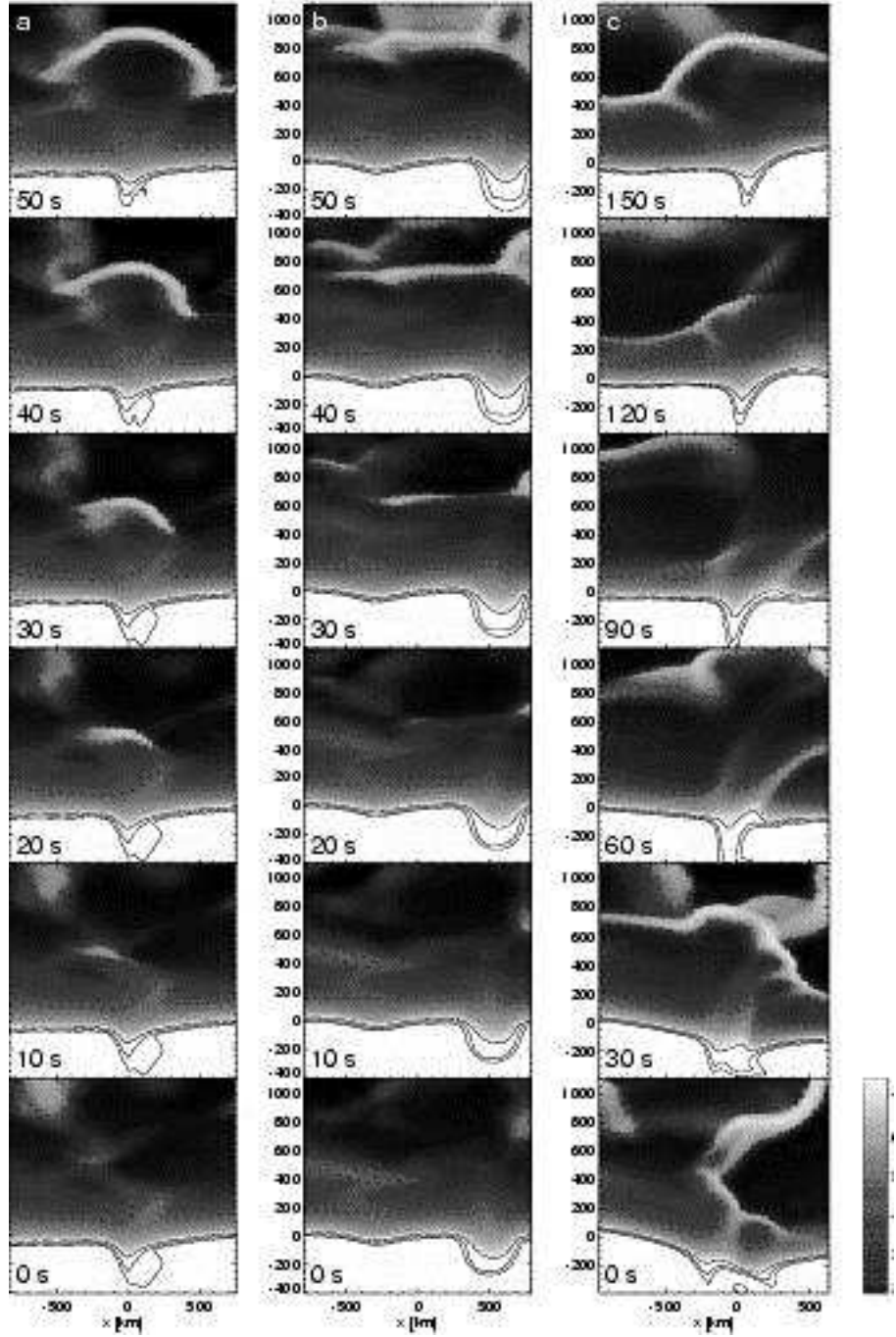}
  \caption{ 
        Formation and propagation of shock fronts:  
        Each column shows a time sequence of vertical slices (temperature) taken  
        from different positions and times of the \mbox{3-D} model.  
        \mbox{{\textbf a)} (left}   column) arch-like/spherical wave,  
        \mbox{{\textbf b)} (middle} column) plane wave,  
        \mbox{{\textbf c)} (right}  column) waves excited by merging downdrafts.  
        Note the different time steps which are quoted in the lower left corners. 
        The temperature is colour-coded for the range $T = 2000$~K to 
        $T = 7500$~K. 
        Additional contour lines are present for 
        $T = 5000$~K (dotted),
        and $T = 7500$~K, 
        $9000$~K, and $10000$~K (all solid).} 
  \label{fig.wavexzseq} 
\end{figure*} 

Acoustic waves are excited by various processes concentrated in the
uppermost layers of the solar convection zone. Excitation processes
have been investigated by means of hydrodynamical modelling by
\citet{skartlien00c}, \citet{nordlund01}, and
\citet{stein01}. Skartlien et al. study the collapse of small granules
which leads to transient wave emission.  Nordlund \&\ Stein focus on
the interaction of convection with resonant oscillatory modes to
derive an estimate of the power input into the solar $5$~min
oscillations. Like the afore mentioned authors, we observe in our
model the excitation of both propagating and standing acoustic
waves. The standing waves are the model analogs to the solar $5$~min
oscillations. Together with the propagating waves they generate a
complex interference pattern in the photospheric and chromospheric
layers, where shocks are frequently formed.

Fig.~\ref{fig.oscheight} illustrates the distribution of power among
radial oscillations as a function of height. Fourier spectra were
calculated for a $151$~min long time sequences of the horizontally
averaged vertical velocity component at each height independently, and
integrated over frequency bands roughly centred around periods of
5~min and 3~min. Figure~\ref{fig.oscheight} shows that the dominant
contribution to the velocity power shifts from the 5-min band to the
3-min band at around $z \sim 1200$~km. We find no significant power in
the low frequency band (periods larger than $\sim 420$~s), while the high
frequency band (periods below $\sim 140$~s) contributes some power in
the higher layers. 

The absolute energy of the oscillatory motions (not shown) decreases
in all bands with increasing height. The largest energies are found in
the deepest layers, indicating that the excitation of the oscillations
takes place in the convection zone for all frequencies. The 5-min band
lies below the acoustic cut-off frequency ($\sim 5.5$~mHz) rendering
these waves evanescent while in the 3-min band some frequencies allow
propagating waves. This implies a stronger damping in the 5-min band,
and explains why the ``3-min'' oscillations dominate in the
chromosphere: the decline of energy with height is more pronounced in
the 5-min band than in the 3-min band. A localised non-linear process
converting oscillatory energy in the 5-min band into energy in the
3-min band is not readily apparent.

\begin{figure}[t] 
  \resizebox{\hsize}{!}{\includegraphics{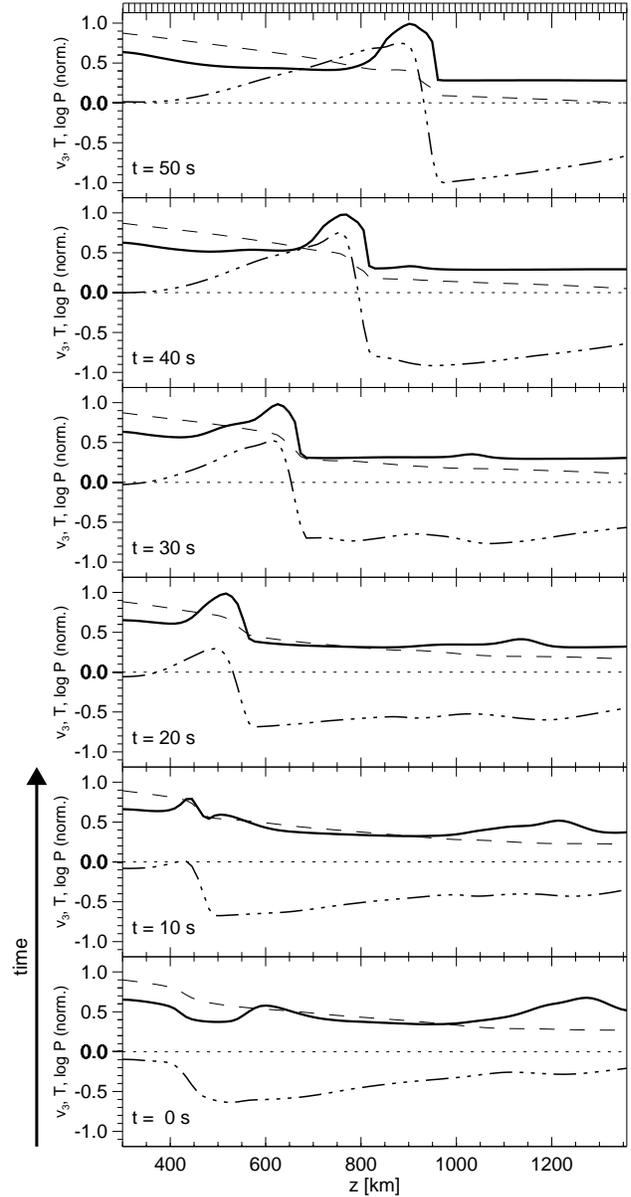}} 
  \caption{Vertical profiles of the flow field shown in  
    \mbox{Fig.~\ref{fig.wavexzseq}.a} at different times along the horizontal  
    position $x = 0$~km. 
    We plot the temperature (solid), the vertical velocity component  
    (triple-dot-dashed), and the logarithmic pressure (dashed) on a linear  
    scale.  
    The data range $0$ to $1$ corresponds to  
    $0$~K to $7000$~K in temperature,  
    $-1.59$ to $4.50$ (cgs units) in the logarithmic pressure, and  
    $0$~$\mathrm{km\,s}^{-1}$ to $15$~$\mathrm{km\,s}^{-1}$ in the vertical  
    velocity component.  
    The vertical grid is shown at the top of the figure.} 
  \label{fig.waveprof} 
\end{figure} 

\begin{figure*}[t] 
  \centering 
  \resizebox{\hsize}{!}{\includegraphics{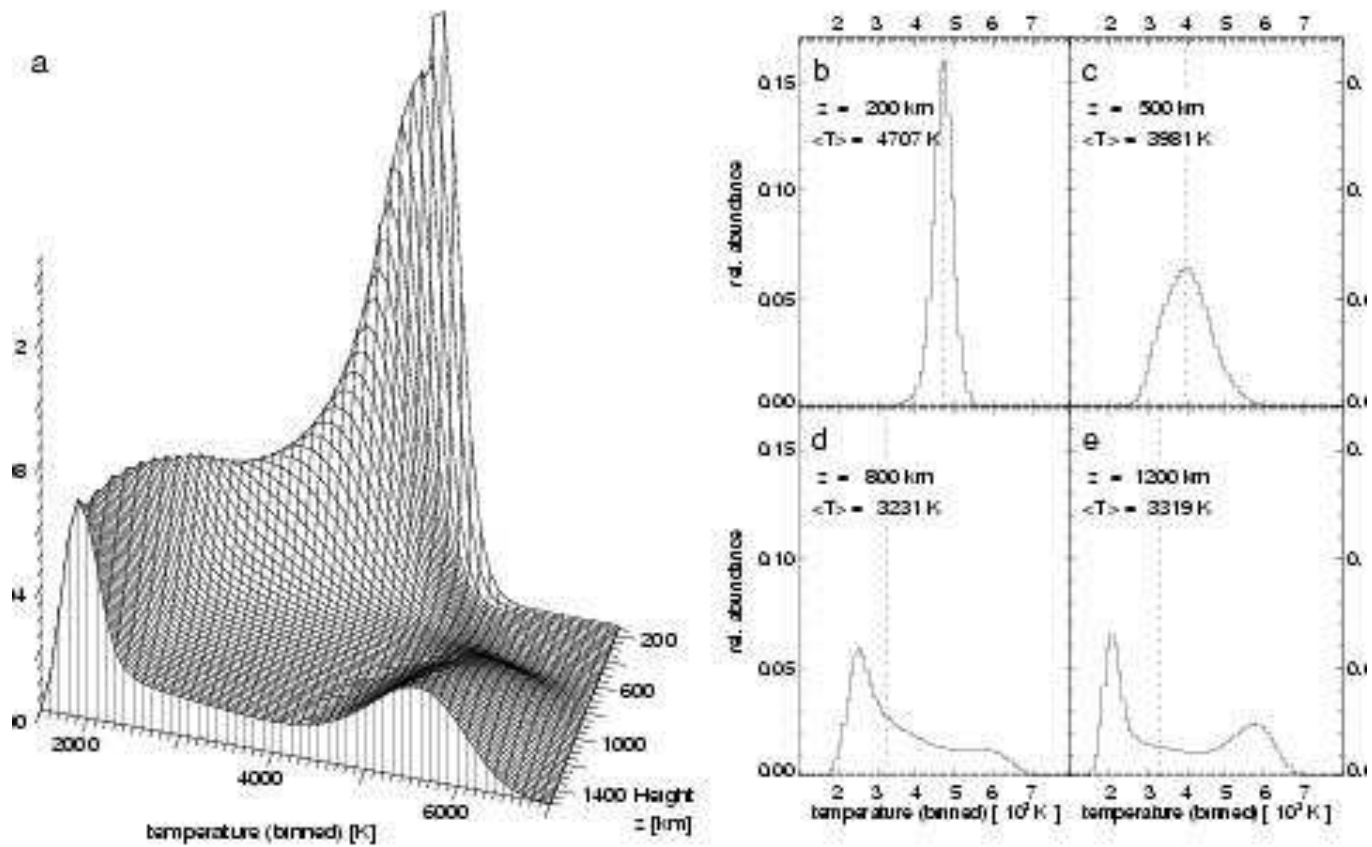}}
  \caption{ 
    Temperature histograms for the \mbox{3-D} model
    ($151$~min of simulation time).     
    For each height step (and all time steps) the temperature values of all grid 
    cells within the  
    corresponding horizontal plane are sorted into temperature bins of  
    $\Delta T = 100\,{\rm K}$.  
    The vertical axis denotes the relative abundance of grid cells within a  
    temperature bin with respect to all cells at that height. 
    \textbf{a)} Height-dependent histogram surface.   
    \textbf{b-c)} Histograms at fixed heights in the photosphere and 
    \textbf{d-e)} in the chromosphere.  
    The dotted lines represent the corresponding mean temperature.} 
  \label{fig.tbifurcation} 
\end{figure*} 

Examples of propagating waves and shock formation are shown in
Fig.~\ref{fig.wavexzseq}.  The example of the left-most column
\mbox{(Fig.~\ref{fig.wavexzseq}.a)} is displayed more quantitatively
in Fig.~\ref{fig.waveprof} for further discussion below.  
It shows the case of a rather localised
shock which was triggered by pressure disturbances emerging from the
downflow region visible in the deeper layers. The formation of a
spherically shaped shock is a frequent pattern. The spherical shock
front appears as an upward travelling arch-like feature in our \mbox{2-D}
cuts.   
The middle column in Fig.~\ref{fig.wavexzseq} shows an example
of a front which is horizontally more extended.  In movies such events
appear often as if the front detaches over a broader area from the
photospheric granulation pattern.  It can extend over more than one
granule and tends to preserve the shape of the granular
pattern for some time. In the simulation, preferentially
resonant modes of long horizontal wavelength are excited. They provide
the horizontally coherent oscillations which are necessary to produce
these extended horizontal wave fronts.  The right-most column
\mbox{(Fig.~\ref{fig.wavexzseq}.c}) shows the formation of shocks
above merging downdrafts, i.e., downflows in the intergranular lanes.
This kind of event corresponds to the collapse of small granules and
has already been investigated in detail by \citet{skartlien00c}.  From
the vertical slices in \mbox{Fig.~\ref{fig.wavexzseq}.c} it can be
seen how two downdrafts are advected horizontally and eventually
merge, producing a stronger and more extended downdraft.  During
the process upward propagating waves are excited which may transform
into shocks in higher layers.  Moreover, a strong downdraft is often
accompanied by shocks of a different nature.  They come about by fast
horizontal flows towards the downdraft.  Shocks form where the flow is
turned into the downdraft. In \mbox{Fig.~\ref{fig.wavexzseq}.c} they
are visible as roughly vertical features attached to the edges of a
downdraft.  These shocks interact with the shocks associated with
the wave field (see frames at $60$~s to $120$~s). Note that
Fig.~\ref{fig.wavexzseq} shows particularly clean examples of the
types of shock events encountered in the simulation. Usually, the
pattern of shocks is very entangled, and often all features discussed
before are present at the same time.

The wave depicted in Fig.~\ref{fig.waveprof} (see also
\mbox{Fig.~\ref{fig.wavexzseq}.a}) is an extreme example as a positive
vertical velocity of $v_z \approx 11$~$\mathrm{km\,s}^{-1}$ is reached
in the chromosphere.  Most velocities are smaller.  We find
approximate upper limits for $95\,\%$ of all upward directed vertical
velocities, depending on height: $\approx 4.9$~$\mathrm{km\,s}^{-1}$
at $z = 800$~km and $\approx 7.0$~$\mathrm{km\,s}^{-1}$ at $z =
1000$~km.  In contrast to one-dimensional simulations, the waves in
our \mbox{3-D}~model do not only propagate in the vertical direction
but also horizontally.  At a height of $z = 1000$~km we find that
$95\,\%$ of all grid cells exhibit horizontal velocities of less than
$\approx 12$~$\mathrm{km\,s}^{-1}$ and $50\,\%$ have values of $\approx
5$~$\mathrm{km\,s}^{-1}$ and below.

An important point is illustrated in Fig.~\ref{fig.waveprof}: shocks
are preferentially formed in low-density material which is flowing
down from above at high velocities. The material has been pushed
upwards by a precursory wave and now falls back again. The shock front
is travelling upstream into the down-flowing material.  In extreme
cases the downflowing material is close to free-fall conditions, and
flow velocities exceed the local sound speed.  The
\mbox{1-D}~simulations by \citet{carlsson97a} exhibit a similar shock
structure (see their Fig.~14). Judging from the same figure, Carlsson
\&\ Stein find typically at most one well developed shock in the
photospheric and chromospheric layers at any given instant in time.
Looking at one particular vertical column in our \mbox{3-D} model we
make a similar observation, finding typically one, sometimes two
fronts.  While in their piston-driven model Carlsson \&\ Stein derive the
wave excitation semi-empirically from observed time sequences of
photospheric oscillations, the shock frequency in our case is a
natural outcome of the simulation. The spatial shock frequency
translates into a temporal recurrence of shocks on a time scale of
$\sim$2\--3~min (see also Fig.~\ref{fig.ttime}). 

\begin{figure*}[t] 
\centering 
  \resizebox{\hsize}{!}{\includegraphics{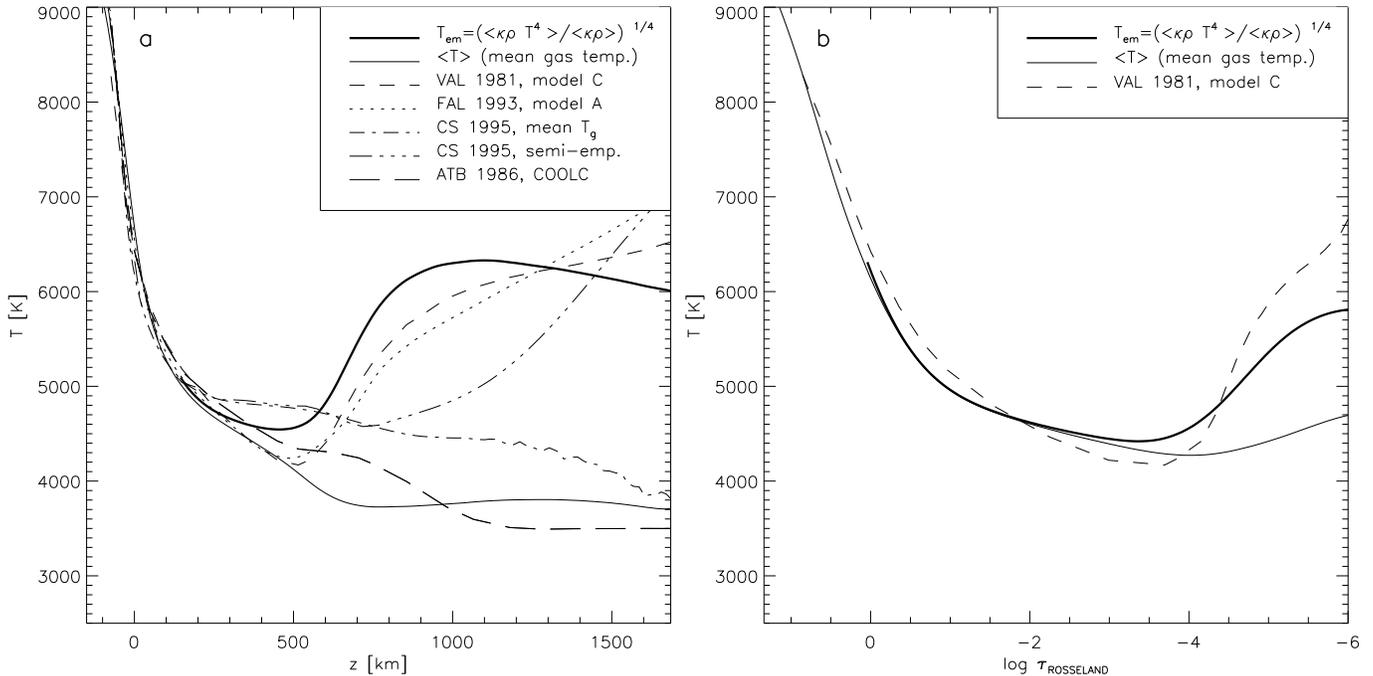}} 
  \caption{Temperature stratifications of different models on a geometric height  
    scale (\textbf{a}) and on an optical depth scale (\textbf{b}): Horizontally  
    and temporally averaged grey emissivity temperature and mean gas temperature  
    for the \mbox{3-D} model, model C by \citet{val81}, model A by  
    \citet{fal93}, mean gas temperature and semi-empirical stratification of the  
    dynamical model by \citet{carlsson95}, and COOLC by \citet{ayres86}.} 
  \label{fig.tavg} 
\end{figure*}

\subsection{Thermal bifurcation} 
\label{sec:tbifurcation} 

Although the chromospheric pattern evolves on very short time scales (see 
Sect.~\ref{sec:struct}), the general picture remains the same in time, i.e., the 
chromosphere appears as a network of hot matter with intermittent cool regions.  
This thermal bifurcation can be quantified via a height-dependent temperature 
histogram.  
For each horizontal slice in the model (constant height~$z$ for each slice) a  
histogram of the temperature values is calculated for temperature bins of  
\mbox{$\Delta T = 100$~K}.  
The result is shown in Fig.~\ref{fig.tbifurcation}.  
In the photosphere, the temperature is distributed close to a mean value with  
only moderate deviations, whereas in the chromosphere, the distribution splits 
up into low and high temperatures.  
Again, this indicates the co-existence of a cool background and hot shocked  
material.

To facilitate a rough comparison with multi-component models  
\citep[e.g.,][]{ayres86,avrett95,ayres96,ayres02},  
we give approximate values for a hot and a cool component of our model  
chromosphere (intermediate values are neglected):  
Above $z = 800$~km the hot temperature ridge in  
Fig.~\ref{fig.tbifurcation} peaks at  
\mbox{$T_\mathrm{hot} = 5500 - 5900$~K},   
whereas the cool temperature peak decreases with height from  
$T_\mathrm{cool} = 2600$~K at $z = 800$~km to  
$\approx 2000$~K for the upper layers of the model chromosphere.  
Thus, the hot component is comparable to the temperatures in the semi-empirical 
models C by VAL and C' by \citet{maltby86} in the height range  
$800 - 1000$~km and $900 - 1100$~km for the model A by FAL,  
respectively.  
The cool component is much colder than COOLC by \citet{ayres86} and COOL0 
by \citet{ayres96}. It is much more like COOL1 \citep{ayres02} around  
$z = 800$~km which, however, is only valid if there is a dominating warm  
component.

\subsection{Temperature stratification} 
\label{sec:tempstrat} 
 
In this section we discuss the consequences of the thermal bifurcation for the  
average temperature stratification.  
The horizontally and temporally averaged gas temperature for the sequence of  
$151$~min simulation time from our model (thin solid line in  
\mbox{Fig.~\ref{fig.tavg}.a}) decreases with height until it reaches values  
between $3800$~K and $3700$~K above $z = 730$~km,  
i.e., in the chromosphere.  
It does not show a notable temperature minimum  nor a significant temperature  
increase in the chromosphere like it is the case in the semi-empirical models by  
VAL and FAL (see \mbox{Fig.~\ref{fig.tavg}.a)}.  
This is qualitatively similar to the mean gas temperature profile in
the \mbox{1-D}~simulation by \citet{carlsson95} which also does not show a
temperature increase (see \mbox{Fig.~\ref{fig.tavg}.a)}.  
However, we obtain chromospheric gas temperatures which are much lower
than in the simulations by CS. In fact, our mean chromospheric gas
temperature lies about $1000$~K below the (grey) radiative equilibrium
temperature of $4680$~K. The mean temperature stratification is
roughly comparable to model COOLC by \citet{ayres86}, which was
constructed as the cool constituent in a multi-component model (see
\mbox{Fig.~\ref{fig.tavg}.a}, where we converted the original column mass
density scale into a geometrical height scale on the basis of model~C
by VAL).

The semi-empirical models are based on spatially and temporally
averaged intensities and thus refer to a static and homogeneous
chromosphere.  We note that the mean gas temperature from our model
matches almost perfectly the semi-empirical models up to a height of
$z\approx 500$~km. Above that height, the thermal bifurcation becomes
increasingly significant, i.e., the temperature fluctuations become
large (see Figs.~\ref{fig.tbifurcation} and
\ref{fig.dtrms}). Clearly, the assumption of spatial 
and temporal homogeneity is not valid in the chromosphere,
and any one-dimensional static description must fail.

CS pointed out that the chromospheric temperature rise in the
semi-empirical models is only an artifact caused by the ``temporal
averaging of the highly nonlinear UV Planck function''.  Furthermore,
CS confirmed this by calculating a temperature distribution for their
dynamical model in a similar way as VAL.  They adjusted a steady-state
temperature stratification to reproduce the time-averaged continuum
and line intensities as a function of wavelength which are a result of
their dynamic simulation.  The semi-empirical model derived in this
way by CS is a much better fit to the models VAL and FAL (see
\mbox{Fig.~\ref{fig.tavg}.a}).

Since no wavelength-dependent intensities are available for our
simulation (except for a few images similar to those shown in 
\mbox{Fig.~\ref{fig.xyslices}.g-h}), 
we calculated a qualitatively similar quantity, namely an
``average grey emissivity temperature'', by averaging the grey
emissivity $\kappa\,\rho\, T^4$, where $\kappa$ is the opacity and
$\rho$ the density.  The corresponding emissivity temperature
$T_{\mathrm{em}}$ is then evaluated as:
\begin{equation} 
T_{\mathrm{em}} (z) =  
\left\langle 
\left({\frac{\langle\,\kappa\, \rho\, T^4\,\rangle_{x, y}} 
{\langle\,\kappa\, \rho\,\rangle_{x, y}}}\right)^{1/4} 
\ \right\rangle_{t}
.
\end{equation} 
The brackets $\langle\,.\rangle_{x, y}$, $\langle\,.\rangle_{t}$
indicate horizontal and temporal averaging, respectively.  The
resulting average temperature profile, calculated on a geometrical
scale \mbox{(thick solid line in Fig.~\ref{fig.tavg}.a)} is indeed
similar to model C by VAL and model A by FAL.  It exhibits a
temperature minimum at approximately the same height; the temperature
values reached in the middle chromosphere are comparable.  This
qualitative match is better than expected from such a crude
approximation.  Thus, like CS we are able to
produce an emissivity temperature stratification qualitatively similar to
the semi-empirical models, without a significant increase in the mean gas 
temperature.

The averages presented so far are calculated on a geometrical height
scale.  In contrast, the average grey emissivity temperature and the
simple arithmetic average shown in \mbox{Fig.~\ref{fig.tavg}.b} are
calculated on the Rosseland optical depth scale which already
incorporates the distribution of opacity and density. Hence, 
the emissivity temperature is given by $T^4_{\tau}$ averaged over 
surfaces of constant optical depth.

We note that the
mean chromospheric gas temperature obtained from averaging on
the optical depth scale are systematically higher (but still below
the radiative equilibrium value) than those on the
geometrical height scale; the minimum values differ by more than
$500$~K.  That is caused by the fact that fluctuations appear much
smaller on surfaces of equal optical depth \citep[see
e.g.,][]{uitenbroek00b}.  In a wave front the optical depth increases
significantly.  Thus, averaging on an optical depth scale is done on
surfaces which are not plane but shaped by the spatial inhomogeneities
while averaging on a geometrical height scale is done on strictly
plane surfaces which cut through the inhomogeneities.  Consequently,
the temperature distribution on a surface for a particular optical
depth differs from the one for a corresponding geometrical height,
leading to different horizontal averages and thus different
temperature stratifications.

\subsection{RMS-temperature fluctuations} 
\label{sec:tfluct} 

\begin{figure}[t] 
  \resizebox{\hsize}{!}{\includegraphics{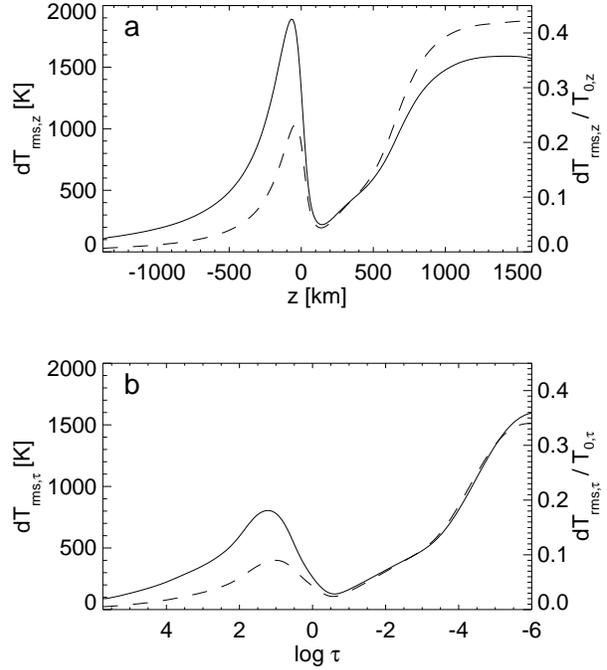}} 
  \caption{Horizontally and temporally averaged temperature fluctuation: 
    Absolute deviations $dT_{\rm rms}$ (solid, left axis) and relative deviations
    $dT_{\rm rms}/T_0$ (dashed, right axis) on the geometrical height scale 
    (\textbf{a}) and on the optical depth scale (\textbf{b}).} 
  \label{fig.dtrms} 
\end{figure} 

\begin{figure}[t] 
  \resizebox{\hsize}{!}{\includegraphics{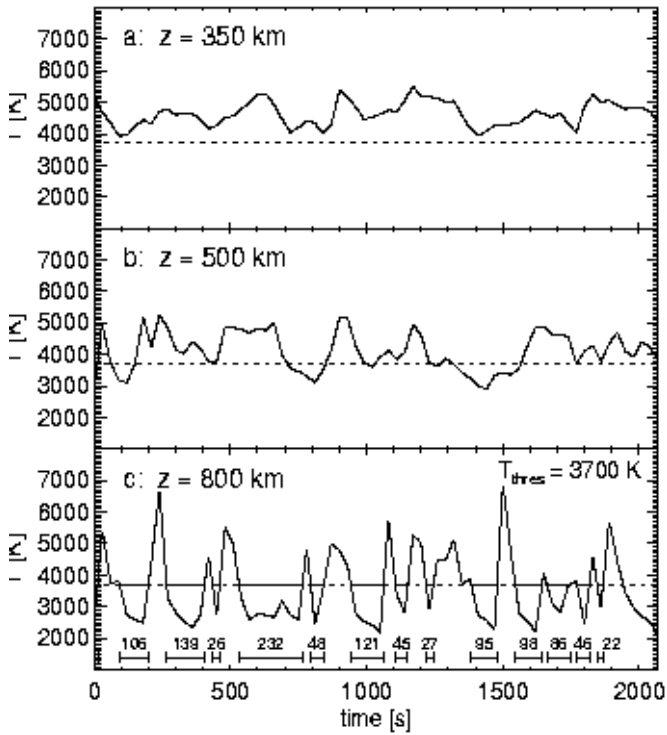}} 
  \caption{Variation of temperature with time for single grid cells at different  
    heights.  
    In panel~\textbf{c)} the cool episodes (time intervals with  
    $T < 3700$~K) are marked with horizontal bars, together
    with the duration in seconds.} 
  \label{fig.ttime} 
\end{figure} 

Here, we quantify the rms-temperature fluctuations which are another measure 
characterising the thermal structure. They are defined by  
\begin{equation} 
  \label{eq.dtrms} 
  \frac{dT_{{\mathrm{rms}}}}{T_{0}} =  
  \frac{\sqrt{\langle(\,T-T_0\,)\,^2\rangle_{x,y,t}}}{T_{0}} 
\end{equation} 
where $T_{0} =\, \langle T \rangle_{x,y,t}$\,  is the temporally and  
horizontally averaged temperature stratification.  
The quantity $dT_{\rm rms}/T_0$ has been calculated on a geometrical and  
on an optical depth scale for the same model sequence as in  
Sect.~\ref{sec:tempstrat} (Fig.~\ref{fig.dtrms}). 
It is strongly height-dependent as can also be seen directly from the
horizontal slices in Fig.~\ref{fig.xyslices} for different heights and
from the temporal temperature variation in Fig.~\ref{fig.ttime}.
Obviously, the lower layers of the solar atmosphere in our model are
relatively homogenous with only small temperature fluctuations, in
contrast to the inhomogeneous chromosphere.
 
Like for the temperature stratification (Sect.~\ref{sec:tempstrat})
there is a difference between the geometrical height scale and the
optical depth scale. Again the temperature deviations are generally
much smaller on a surface of a particular optical depth than for a
corresponding geometrical height \citep[see, e.g.,][]{uitenbroek00b}.
In both cases the average lies below $dT_{\rm rms}/T_0 \approx 0.42$.
For particular vertical positions and time steps, maximum values of 
$\approx 1.0$ can be reached.
 
A comparable quantity $\delta T / T$ has been used by
\citet{kalkofen01} to distinguish between the two opposing cases of a
hot chromosphere with small temperature fluctuations \mbox{($\delta
T/T \approx 0.1$)} and a cool one with large fluctuations
\mbox{($\delta T/T \approx 10$)}. Our model lies in between these
cases. As mentioned earlier, the inclusion of time-dependent
ionisation likely leads to higher temperature peaks and accordingly to
larger temperature deviations.

\subsection{Cool regions} 
\label{sec:coolregion} 

\begin{figure}[t] 
  \resizebox{\hsize}{!}{\includegraphics{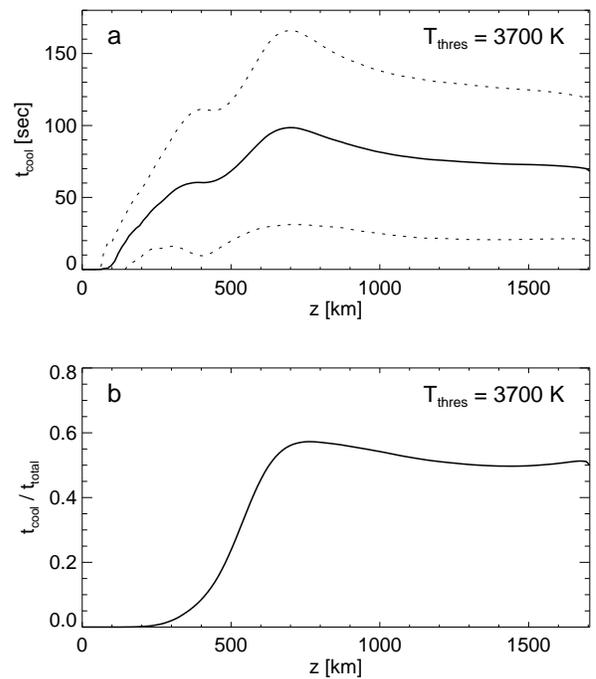}} 
  \caption{Height-dependent duration of cool episodes for a threshold  
    temperature of $T_{\rm thres} = 3700$~K in the \mbox{3-D}~model.   
    {\textbf a)}~Absolute values for the average (solid) and the average plus/minus  
    standard deviation (dotted); {\textbf b)}~Ratio of integrated cool time to total time.} 
  \label{fig.tcool} 
\end{figure} 

\begin{figure}[t] 
  \resizebox{\hsize}{!}{\includegraphics{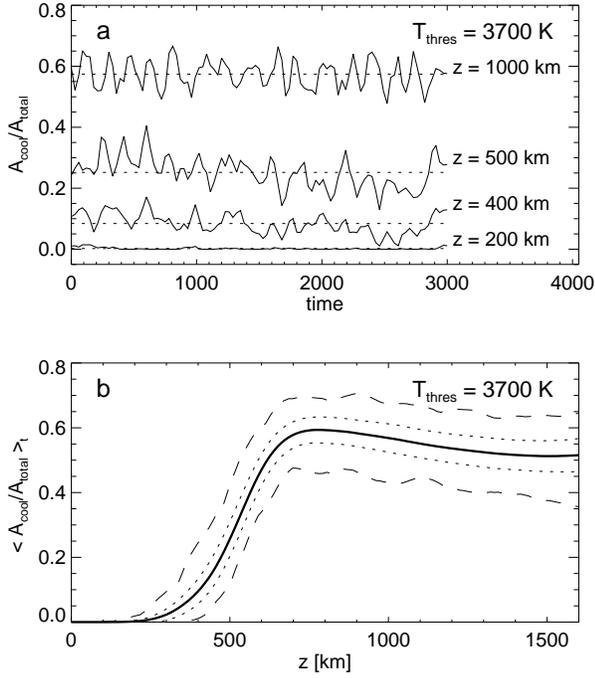}} 
  \caption{Ratio of cool area to total area for a threshold temperature of  
    $T_{\rm thres} = 3700$~K in the \mbox{3-D}~model.   
    {\textbf a)}~Variation with time (solid) and  
    time-averages (dotted) at different heights; 
    {\textbf b)}~Height-dependent time-average (solid),  
    $\pm$ standard deviation (dotted), and maximum and minimum values (dashed).} 
  \label{fig.cregarea} 
\end{figure} 

As a consequence of the propagating shock waves, the temperature at a fixed  
position in the model chromosphere varies by several $1000$~K  
with time, featuring sharp temperature peaks on top of a cool background  
(\mbox{Fig.~\ref{fig.ttime}.c}).   
Observations of the infrared CO fundamental vibration-rotation lines imply  
temperatures as low as $3700$~K \citep[e.g,][]{uitenbroek00a} which  
also represent a lower limit for the average temperature stratification of the  
\mbox{3-D}~model (see Sect.~\ref{sec:tempstrat}).  
We adopt this temperature as a threshold value and determine how long the  
temperature at a fixed position in the model stays below this value.  
In the following we will refer to these time intervals with  
$T < T_{\rm thres} = 3700$~K as cool episodes. 
In \mbox{Fig.~\ref{fig.ttime}.c} such episodes are illustrated.  
The duration of a cool episode is influenced by the local background temperature  
and the temperature fluctuations due to the propagating waves. 
Therefore, it depends on height.  
However, the average duration stays more or less constant throughout a wide  
height range in the chromosphere.  
For the \mbox{3-D} model, we determined the average duration of the cool  
episodes in the chromosphere to be $70 - 100$~s   
\mbox{(Fig.~\ref{fig.tcool}.a)}.   
In some cases the cool episodes are much longer, up to several hundred 
seconds.

With regard to a more global view of the chromosphere not only the
duration of single cool episodes but also the sum of all durations is
interesting.  The temperature at a fixed position in the chromosphere
of the \mbox{3-D}~model stays roughly half of the time below $T_{\rm
thres} = 3700$~K (Fig.~\ref{fig.tcool}.b).  In the lower photosphere
cool episodes are rare and thus negligible with regard to the total
time.

The spatial scales of the cool regions might also be  
interesting for the interpretation of observations.  
The average radius of a cool region is hard to determine because the regions are  
often not closed structures like a cloud but are connected to other cool regions  
in a complicated way.  
As can be seen from Fig.~\ref{fig.xyslices} the spatial scales are on average  
comparable to the granulation, except for some rare cases with larger cool  
areas.  
The fraction of the integrated cool area at a particular height shows only  
relatively small temporal fluctuations (Fig.~\ref{fig.cregarea}.a).  
Thus, the model chromosphere is never completely cold and never completely hot.  
There are always cool regions next to a hot component.  
The height-dependent time-average of the cool area fraction (see  
Fig.~\ref{fig.cregarea}.b) is equal to the average ratio of cool time to total 
time (Fig.~\ref{fig.tcool}.b) because both represent the horizontally and 
temporally averaged number of grid cells with temperatures below the threshold 
value.   
On average $50$ to $60\,\%$ of the whole time and of the whole area in a 
horizontal slice of the model chromosphere has a temperature below $3700$~K.  
Consequently, this cool component is not just a minor constituent 
in our \mbox{3-D} model.

\subsection{Carbon monoxide} 
\label{sec:carbonmon} 

\begin{figure*}[t] 
  \resizebox{\hsize}{!}{\includegraphics{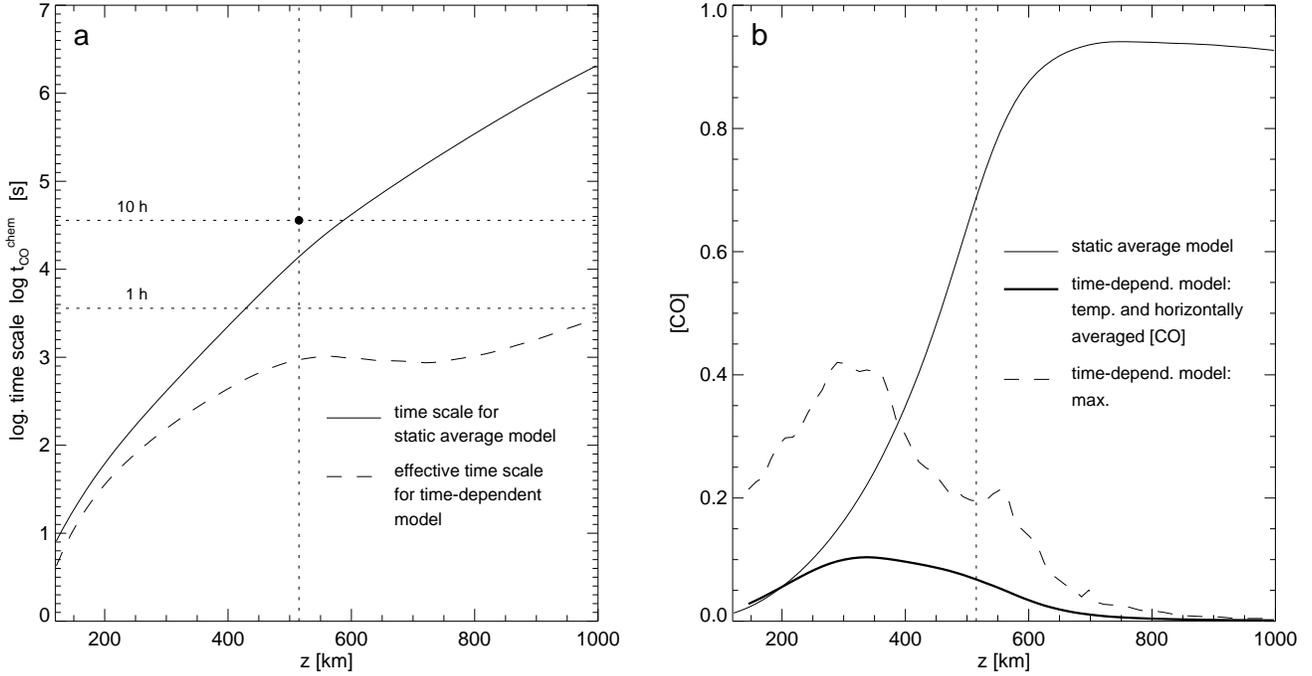}} 
  \caption{Formation and destruction of carbon monoxide.  
    {\textbf a)}~Chemical time scale  $t_{\element{CO}}^{\rm chem}$ for the  
    averaged \mbox{3-D} stratification (solid) and for the time-dependent  
    calculation  $t_{\element{CO}}^{\rm chem, dyn}$ (dashed).  
    The black dot indicates a time scale of $10$~hours at the classical  
    temperature minimum \citep[][]{ayres96}.  
    {\textbf b)}~Equilibrium CO concentration for the averaged  
    \mbox{3-D}~stratification (thin solid), and average CO concentration of 
    the time-dependent calculation (thick solid) together with the 
    corresponding maximum and minimum values (dashed). 
    $[\element{CO}] = 1$ means all carbon is bound in CO~molecules.} 
  \label{fig.co} 
\end{figure*} 

It is not obvious how the variable hydrodynamic conditions affect the
formation, dissociation, and spatial distribution of CO molecules in
the outer solar layers. 
Here we present the results of a simple
time-dependent calculation of the CO concentration, demonstrating that
the predicted height distribution of CO can be very different in a
static and in a dynamic solar atmosphere, because the reaction rates
are highly non-linear functions of temperature.

For simplicity, we assume that CO is formed by direct radiative  
association, $\element{C} + \element{O} \rightarrow \element{CO} + h\nu$,
and is destroyed by collisional dissociation, 
$\element{CO} + \element{H} \rightarrow \element{C} + \element{O} + 
\element{H}$. 
In this case, the temporal evolution of the CO concentration
$[$\element{CO}$]$ is governed by the differential equation
\begin{equation} 
  \frac{{\rm d}}{{\rm d}t} \left([\element{CO}] \right) = 
   k_1 - k_2\,[\element{CO}],
  \label{eq.co1} 
\end{equation} 
where $[\element{CO}] = n_{\element{CO}}/(n_{\element{C}} + n_{\element{CO}})$; 
a value of 1 means that all carbon is bound in CO molecules.
According to \citet{ayres96},
the constants $k_1$ and $k_2$ depend on the number density of neutral
hydrogen, $n_{\element{H}}$, and on the temperature $T$ as
\begin{equation} 
  k_1 = 2.5 \, 10^{-5} \, n_{15}\, \tilde{T}^{0.6} 
  \label{eq.k1} 
\end{equation} 
and 
\begin{equation} 
  k_2 = k_1 \, \left(1 + 40\, \tilde{T}^{22.2}\right)  
  \label{eq.k2} 
\end{equation} 
with the notations $n_{15} = n_{\element{H}}/(10^{15}$~cm$^{-3}$) and  
$\tilde{T}=T/(5000$~K).  

In a static environment, the equilibrium CO concentration,
\begin{equation} 
  [\element{CO}]_{\rm eq} = \frac{k_1}{k_2} =  
  \left(1 + 40\,\tilde{T}^{22.2}\right)^{-1},
  \label{eq.coeq} 
\end{equation} 
is approached  with a characteristic time scale
\begin{equation} 
  t_{\element{CO}}^{\rm chem} = k_2^{-1} = \frac{4\, 10^4}{n_{15}} \,  
  \frac{\tilde{T}^{-0.6}}{1 + 40\,\tilde{T}^{22.2}}\;\; [{\rm s}]. 
  \label{eq.cotchem} 
\end{equation} 
The characteristic time scale $t_{\element{CO}}^{\rm chem}$ according
to Eq.~(\ref{eq.cotchem}) and the equilibrium CO concentration
$[\element{CO}]_{\rm eq}$ according to Eq.~(\ref{eq.coeq}) are plotted
as a function of height for the mean temperature and density structure
of our \mbox{3-D} simulation in \mbox{Fig.~\ref{fig.co}} (thin solid
lines). $t_{\element{CO}}^{\rm chem}$ varies by orders of magnitude,
from $\approx 0.1$~s at $z = 0$~km ($\tau \approx 1$) to
$\ga 10^6$~s at $z = 1000$~km. This variation is
partly due to the temperature dependence of $t_{\element{CO}}^{\rm
chem}$, but mainly due to the density factor. In the lower
chromosphere ($z \ga 600$~km), $[\element{CO}] \ga 0.9$,
implying that almost all carbon in the chromosphere is bound in CO.

The situation is quite different in a dynamic atmosphere. We have
investigated the time-dependent case quantitatively, using $196$ 
representative vertical columns of the \mbox{3-D} model described
before.  The simulations then provide the temperature and density
variations at each point for the time interval of $151$~min. 
These
prescribed fluctuations translate into time-dependent coefficients
$k_1$ and $k_2$ according to Eqns.~(\ref{eq.k1}) and (\ref{eq.k2}).
We have solved Eq.~(\ref{eq.co1}) with these time-dependent
coefficients for each point in the selected columns, using a standard
fourth-order Runge-Kutta method.  If necessary, the time-sequences
from the simulations are repeated until a (dynamic) equilibrium is
obtained.
 
The resulting horizontally and temporally averaged CO concentration is
shown in \mbox{Fig.~\ref{fig.co}.b} (thick solid line). It differs
dramatically from the distribution found in the static mean atmosphere
(thin solid line), except for the deep photosphere ($z \la
200$~km). In the dynamical atmosphere, CO is present in the
photosphere and in the low chromosphere with a maximum concentration
of only $\langle [\element{CO}] \rangle_t \approx 0.10$ at $z \approx
340$~km. Very little CO is found in the layers above $z \approx
700$~km.

This finding is in line with the calculations by \citet{asensio03}
which are based on the \mbox{1-D} numerical simulations by CS. 
They, too, state that no significant CO concentration should be present at 
heights greater than $\approx 700$~km. 
Furthermore, the difference in the CO concentration between the 
static and the dynamic approach becomes larger with increasing height 
\citep[see Fig.~2 in][]{asensio03} which is qualitatively similar to our results 
(see \mbox{Fig.~\ref{fig.co}.b}).  
Hence, we agree with \citeauthor{asensio03} that detailed nonequilibrium CO 
chemistry must be taken into account.

The reason for the difference between the static and the dynamic model
presented in this work, however, is 
related to the fact that CO is rather efficiently
destroyed during the passage of high-temperature regions (shock
fronts), where the chemical reaction time scales are short. In the
subsequent cool phases (see also Sect.~\ref{sec:coolregion}), reaction
time scales are much longer, so the CO concentration builds up rather
slowly and reaches only moderate levels before the next
high-temperature event occurs. The low CO concentration in the upper
atmosphere is thus a consequence of the onset of shock formation and
the related higher temperature peaks 
(see \mbox{Fig.~\ref{fig.tbifurcation}.a}).

If chemical time scales are short compared to the hydrodynamical time
scales, $t_{\element{CO}}^{\rm chem} \ll t_{\rm HD}$, then chemical
equilibrium is reached instantaneously, and $[\element{CO}]_{\rm
eq}^{\rm dyn} \approx \langle k_1 / k_2 \rangle_t$.  Hence, we find
that the CO concentration in the lower photosphere ($z \la
200$~km) is reasonably well represented by this
approximation. In these layers, the spatial CO distribution is
tightly correlated with the local temperature: the coolest regions
have the highest CO concentration.

For the higher layers ($z \ga 200$~km), which are of more
interest in this investigation, we find $t_{\element{CO}}^{\rm chem} >
t_{\rm HD}$. The CO concentration is then well approximated by
$[\element{CO}]_{\rm eq}^{\rm dyn} \approx {\langle k_1 \rangle_t}/
{\langle k_2 \rangle_t}$. Since $k_2$ is a highly non-linear function
of $T$, the high-temperature events vastly dominate the time average
and the resulting CO concentration is much smaller than implied by the
mean temperature.

We can also conclude that the dynamical equilibrium CO concentration
is attained on a characteristic time scale
$t_{\element{CO}}^{\rm chem,dyn} = \langle k_2 \rangle_t^{-1}$, 
which is also shown in \mbox{Fig.~\ref{fig.co}.a} (dashed).
The correlation between $[\element{CO}]$ and $T$ is expected to be
poor in the higher layers: the highest concentrations build up in
places with the longest history of relatively undisturbed conditions
($[\element{CO}] \approx 0.4$), and almost no CO is found just behind 
strong shock fronts.

The results described above can only be a first estimate of the height
profile of $[\element{CO}]$ under time dependent conditions, because
the underlying calculations still have severe limitations. More secure
conclusions about the CO distribution in the upper solar atmosphere
have to wait for more detailed future simulations taking into account
(i) the transport of \element{CO} molecules with the flow, (ii) a
more complete chemical reaction network including multi-step reactions
affecting the CO balance \citep[see][]{ayres96,asensio03}, 
and (iii) the back
reaction of the CO concentration on the radiative cooling rate.

\section{Discussion}
\label{sec:discuss}

Quantitatively, the results presented in the preceding sections must
be considered as preliminary, since the physics of 
\mbox{\textsf{CO$^\mathsf{5}$BOLD}}\ are not yet
properly adapted to chromospheric conditions.  In particular, the
assumption of LTE is a poor
approximation in the chromosphere \citep[][]{carlsson02,rammacher03}. 
A realistic treatment should account for deviations
from LTE and also requires the time-dependent computation of
the ionisation of hydrogen and other important species. 
Nevertheless, we believe that some
of the basic features seen in our model are insensitive to the
detailed treatment of thermodynamics and radiative transfer.

The \mbox{3-D} topology of the small-scale chromospheric network we
discovered in our simulation, and its spatial and temporal scales are
expected to be a robust feature.  This is confirmed by test
calculations with different values of the tensor viscosity, a
different grey opacity table, and even with a frequency-dependent
(multi-group) radiative transfer scheme using five opacity bins: the
dynamical properties of these models (like the height-dependent
amplitude of the velocity fluctuations) turn out to be quite
insusceptible to changes of the analysed numerical parameters.
We attribute this to the fact that the dynamics of the model chromosphere  
are governed by the lower layers where the excitation of acoustic waves 
takes place and that the numerical modelling of these layers, i.e., the 
photosphere and the top of the convection zone, is quite realistic. 
In this context, it is reassuring to find prominent chromospheric 
oscillations in the 3-min range whose properties are largely independent of the
numerical details of the simulation.  
The qualitative similarity to observations indicates that the dynamics are indeed 
modelled reasonably well.

The horizontal structure of our model chromosphere, i.e., its topology, 
is reminiscent of observed patterns like the chromospheric ``background
pattern'' found by \citep{krijger01} and the structure of 
\ion{Ca}{ii}~H observations \citep{suetterlin03}. 
The latter, for instance, exhibits spatial scales which 
are comparable to the granulation and thus to the scales which are found 
in our numerical simulation. 
However, the observed patterns originate predominantly from lower layers 
and should therefore not be confused with the patterning of the model 
chromosphere.
The differences might be revealed by determining the time scales on which the 
different patterns evolve. This issue needs to be investigated more 
properly in the future.

In contrast to the spatial scales and the topology of the atmospheric 
patterns, the amplitude of the temperature fluctuations 
in the model chromosphere is more susceptible to the treatment of
radiative transfer. Indeed, the temperature fluctuations are
significantly smaller in the aforementioned test calculation using a
frequency-dependent radiative transfer scheme.  However, we recall
that, up to the mid-chromospheric layers ($z=1000$~km), the peak shock
temperatures in our grey simulation are very similar to those
found by \citet[][]{carlsson94,carlsson97a} and by
\citet{skartlien98}. 
More precisely, in the lower and mid chromospheric regions CS find peak temperatures
which lie only somewhat ($\sim$1000~K) above our values. 

The shock peak temperatures are of importance since many spectral features
are biased towards high temperatures. 
Theoretically, the peak
temperatures depend on the shock strength and are given by the
Rankine-Hugoniot jump conditions. In the absence of radiation, a
conservative numerical scheme guarantees that the jump conditions are
fulfilled, i.e.,\ the post-shock temperature is independent of the
spatial resolution. If radiation is important, however, the peak
post-shock temperature is reduced relative to the theoretical value by
an amount that depends on the spatial resolution of the numerical
grid. This is because the shock heating is stretched out over a finite
time interval, given by the time a volume element needs to cross the
shock front which is smeared out over a number of grid points. In this
case, radiative cooling can reduce the attainable peak temperature if
the radiative cooling time is comparable or smaller than the time
scale of shock heating.  Furthermore, the overall energy dissipation
in the shock is altered due to a change of the effective adiabatic
exponent of the gas.

In our model based on grey radiative transfer all chromospheric layers
are optically thin. Here, radiative cooling times are independent of
the flow geometry and mainly dependent on temperature. The cooling
time at a temperature of 7000~K --- about the highest temperature we
observe in the simulation --- amounts to $\sim$200~s, and is
increasing rapidly for lower temperatures. The dissipation time scale
in the shocks is in the order of a few seconds. This means that the
thermal structure of our shocks is hardly affected by radiation and
primarily given by the shock strength. Only in the most extreme cases
we expect some limiting influence of radiative cooling on the
post-shock temperature. Similarly, the thermal structure of the
post-shock regions is mainly controlled by cooling via adiabatic
expansion.

As mentioned above, the differences in the peak temperatures of our model 
and the simulation by CS  become larger in the higher layers. 
First, CS employ an adaptive grid in their simulation 
with a grid spacing of typically 200~m near shocks which is thus 
much finer than our fixed (vertical) spacing of 12~km.
Furthermore, it appears plausible that our radiative transfer -- based on
Rosseland opacities including lines as true absorption -- produces 
shorter radiative cooling times compared to CS. 
The higher
resolution and longer radiative cooling times in the model of CS
lead us to expect that their shock peak temperatures are also largely
unaffected by radiation.

Two effects can explain the somewhat higher shock temperatures of
CS. 
First, the shock strength in the CS model might simply be higher than in
our case. This could be related to the semi-empirical piston velocity
CS feed in at the bottom of their model, or their 1-D geometry forcing
shocks to remain plane-parallel. As we have seen above, extended
horizontal shocks are more the exception than the rule in our 
\mbox{3-D} simulation; most shock fronts weaken as they propagate
radially away from their source. Another effect is related to our
assumption that thermodynamic equilibrium conditions prevail in the
chromosphere. In a recent paper, \citet{carlsson02} demonstrated that
this is a poor approximation. Ionisation equilibria cannot follow the
rapid thermodynamic changes introduced by the flow. 
One consequence is that the energy which is dissipated in shocks cannot go into
ionisation but has to go into a temperature increase of the post-shock
gas. 
Moreover, accounting for finite recombination time scales instead 
of assuming ionisation equilibrium could reduce the ability of the 
post-shock gas to cool, thus leading to even higher temperatures.  
Since CS account for the thermodynamic non-equilibrium effects,
their shock temperatures should be higher.

We cannot decide on the basis of the available information which is
the reason for the differences in the peak temperatures of the
shocks. However, we conclude that the differences up to the
mid-chromospheric layers ($z=1000$~km) are modest: our peak temperatures
are $\approx 7000$~K compared to $\approx 8000$~K in the CS model.  We
further note that the chromospheric peak temperatures found by
\citet{skartlien98} (see his Fig.~9) are $\la 7000$~K in their case of
frequency-dependent radiative transfer accounting for line scattering,
which is surprisingly close to our result obtained with our grey LTE
radiative transfer.

This supports our conclusion that our grey
radiative transfer employed in this work is more realistic than the
frequency-dependent method available for \mbox{\textsf{CO$^\mathsf{5}$BOLD}}.
The latter method strongly overestimates the (LTE)
cooling in the strong spectral lines (treated as true absorption) and
thus wrongly reduces the maximum attainable temperatures. 
However, note that also the grey radiative transfer is not appropriate 
for chromospheric conditions. 
Rather, a detailed frequency-dependent non-LTE radiative transfer is 
necessary. 
Furthermore, for a
quantitative comparison with the observations it would be necessary to
perform three-dimensional spectrum synthesis, which is planned for the
future.

In contrast to the peak temperatures, the mean chromospheric
temperature (and also the minimum temperatures) in our simulation are
significantly lower than those found by CS (see Fig.~\ref{fig.tavg}),
and also somewhat cooler than in the \citet{skartlien98} model.
Obviously, the mean temperature structure is more strongly influenced
by the treatment of radiative transfer
than it is the case for the peak temperatures.
It should therefore be considered as uncertain. 
Somewhat surprisingly, however, we note that our grey and
our frequency-dependent simulations produce almost identical mean
chromospheric temperature structures.

Nevertheless, the differences in the average temperature stratification 
between the one-dimensional simulation by CS and the presented 
\mbox{3-D}~model can be  understood if one watches the velocity field 
of a region which just has been traversed by a strong shock wave. 
The flows are mostly directed outwards away from the centre of such a region. 
We interpret this as fast and thus adiabatic expansion of the traversed region. 
This "dynamic cooling" is obviously more efficient in \mbox{3-D} than 
\mbox{1-D} simply due to the additional spatial dimensions. 
This effect thus produces lower average and minimum temperatures in our 
simulations compared to those of CS.

Furthermore, 
we point out that the thermal bifurcation in our \mbox{3-D}
model (see Sect.~\ref{sec:tbifurcation}) is not due to the
action of carbon monoxide as a cooling agent. Rather, it is caused by
the acoustic wave field and the resulting dynamic cooling of
adiabatically expanding regions as discussed above. 
Carbon monoxide is only taken
into account in the grey opacity tables so far, and so its real
influence is underestimated.  A similar simulation with a different
grey opacity table without molecular contributions (based on ATLAS6,
\citealt{kurucz70})
leads to very similar results.  We thus conclude that CO plays no
active role in our present simulations. On the other hand, the
calculations described in Sect.~\ref{sec:carbonmon} demonstrate that
there is a non-negligible amount of CO present in the lower
chromosphere.  Hence, a full treatment of CO as a cooling agent might
even amplify the thermal bifurcation of the chromosphere \citep[see, 
e.g.,][]{ayres81,anderson89b,steffen88}.

Although the mean chromospheric temperature of our simulation lies
considerably below the radiative equilibrium temperature,
we find a net radiative cooling of the chromospheric layers: $\langle
\nabla\cdot\vec{F}_{\rm rad}\rangle_t$ is positive here.  This
apparent contradiction can be explained by the presence of
sufficiently strong temperature fluctuations and the highly non-linear
temperature dependence of the radiative heating/cooling rates.
We do not claim, however, that this situation is actually realized in
the solar chromosphere.

Our simulations indicate that the wave generation is mainly controlled
by the large-scale dynamical evolution of the granulation pattern (see
Sect.~\ref{sec:shockwaves}). This is at variance with the classical
picture of the Lighthill-Stein theory
\citep[][]{lighthill52,stein67,stein68} where small-scale turbulent
eddies make the main contribution to the acoustic energy flux. Applying 
the Lighthill-Stein theory to the Sun, \citet[][]{musielak94} find
that the acoustic flux spectrum shows a maximum near $\nu \ga 15$~mHz,
and hence is dominated by ``short period waves''.

The presented simulation can marginally resolve turbulent eddies in the
convection zone with wavenumbers up to $k_{\rm max}
\approx 2\pi/(5 \Delta x)$. According to the classical theory, eddies 
with wavenumber $k$ mostly contribute to the acoustic wave spectrum at
frequency $\omega = k u_{k}$, where $u_{k}$ is the turbulent velocity
of eddies with wavenumber $k$. Since $u_{k} \la 1$~km/s, the
simulation cannot describe the turbulence spectrum beyond $\omega_{\rm
max} \la 30$~mHz, $\nu_{\rm max} \la 5$~mHz. We conclude that our
present model cannot resolve the small-scale turbulence which is
responsible for the sound generation in the Lighthill-Stein theory.
The acoustic flux resulting from our simulation decreases
monotonically with frequency, and so has little in common with the
spectrum predicted by the Lighthill-Stein spectrum. Hence, it appears
doubtful whether the classical theory, based on the assumption of
isothermal, homogeneous, and isotropic turbulence, captures the
essential physics of the violent, highly anisotropic layers at the top
of real stellar convection zones.  We argue that our numerical
simulation correctly represents the basic mode of wave generation,
even at the present spatial resolution.

\section{Conclusions} 
\label{sec:conclusion} 

Based on a detailed \mbox{3-D} simulation of the solar granulation and
the overlying atmosphere, we have studied the generation of waves by
the time-dependent convective flow, and the wave propagation and
dissipation in the higher layers. 
The most important improvements compared to previous numerical simulations 
are (i) self-consistent dynamics without a need for a driving piston like 
done by CS and (ii) a high spatial resolution which is obviously necessary 
for modelling the small-scale structure of the solar chromosphere.   
On the other hand, the LTE treatment of the
thermodynamics and the radiative transfer
is certainly unrealistic in the chromospheric layers. 
We have presented evidence that some of the
basic features seen in our model are nevertheless representative of
the (non-magnetic) internetwork regions of the solar chromosphere.

The main result of the present investigation is the discovery of a
complex network of hot filaments pervading the otherwise cool
chromospheric layers.  
Caused by interaction of standing and propagating hydrodynamic waves
of large amplitude, 
the model chromosphere is a highly dynamical, spatially and temporally
intermittent phenomenon. Its temperature structure is characterised by
a thermal bifurcation: hot and cool regions co-exist side by
side. Temperatures in the hot filaments are high enough to produce
chromospheric emission lines, and the cool ``bubbles'' are cold enough
to form molecular features. Thus, the chromosphere is {\it hot and
cold} at the same time.  This picture of the \mbox{3-D} structure of
the solar chromosphere has the potential to explain the apparently
contradictory observational diagnostics which cannot be understood in
the framework of \mbox{one-dimensional} theoretical or semi-empirical
models.

The presence of strong spatial and temporal temperature fluctuations
has a remarkable consequence: the temperature minimum and the outward
directed temperature rise inferred from semi-empirical models might be
artifacts in the sense that they do not necessarily imply an increase
of the average gas temperature with height.  Our model suggests that
the radiative emission can be sustained by the hot propagating shock
waves even though the main fraction of the chromospheric layers is
cool and the mean gas temperature profile shows an almost monotonic decrease
\-- a conclusion already reached  by
\citet{carlsson94,carlsson95,carlsson97a} on the basis of
\mbox{one-dimensional} hydrodynamical simulations.

We conclude that improved \mbox{3-D} radiation hydrodynamic 
simulations of the kind presented in this work are likely to lead the
way towards a consistent physical model of the thermal structure and dynamics
of the non-magnetic solar chromosphere which eventually can
explain the various observational diagnostics.

\begin{acknowledgements} 
We are grateful to M.~Carlsson, R.~F.~Stein, M.~Wunnenberg, and 
P.~{S{\" u}tterlin} for providing data for comparison and to 
R.~J.~Rutten for advice and critical comments.  
The numerical simulations were carried out on the CRAY~SV1 of the  
{\em Rechenzentrum der Universit\"at Kiel}.   
SW was  
supported by the {\em Deutsche Forschungs\-gemein\-schaft (DFG)}, 
project Ho596/39. 
HGL acknowledges financial support of the {\em Walter Gyllenberg Foundation}  
(Lund), and the Swedish {\em Vetenskapsr{\aa}det}. 
BF is supported by a grant from the Swedish {\em Sch\"onbergs Donation}.
\end{acknowledgements}

\bibliographystyle{aa} 
\bibliography{H4284}

\end{document}